\documentclass[aps,prl,twocolumn,reprint,superscriptaddress,citeautoscript]{revtex4-2}

\usepackage{physics} 
\usepackage{graphicx}
\usepackage{amsmath,amssymb,bm}
\usepackage[normalem]{ulem}

\usepackage{color}

\usepackage{hyperref}
\hypersetup{
	citecolor = blue,
	colorlinks = true,
	urlcolor = blue
}

\newcommand\A{\mathcal{A}}

\renewcommand\S{\mathcal{S}}
\newcommand\T{\mathcal{T}}
\newcommand\N{\mathcal{N}}
\renewcommand\H{\mathcal{H}}
\newcommand\I{\mathcal{I}}

\newcommand\rmT{\mathrm{T}}
\newcommand\rmR{\mathrm{R}}
\newcommand\rmL{\mathrm{L}}

\renewcommand\k{{\bm k}}
\newcommand\p{{\bm p}}
\newcommand\q{{\bm q}}
\renewcommand\r{{\bm r}}
\newcommand\R{{\bm R}}

\renewcommand\d{\partial}
\newcommand\+{\dagger}
\newcommand\<{\langle}
\renewcommand\>{\rangle}

\begin{document}
\title{Thermomagnetic Anomalies by Magnonic Criticality in Ultracold Atomic Transport}

\author{Yuta Sekino}
\affiliation{RIKEN Cluster for Pioneering Research (CPR), Astrophysical Big Bang Laboratory (ABBL), Wako, Saitama, 351-0198 Japan}
\affiliation{Interdisciplinary Theoretical and Mathematical Sciences Program (iTHEMS), RIKEN, Wako, Saitama 351-0198, Japan}
\affiliation{Nonequilibrium Quantum Statistical Mechanics RIKEN Hakubi Research Team, RIKEN Cluster for Pioneering Research (CPR), Wako, Saitama 351-0198, Japan}

\author{Yuya Ominato}
\affiliation{Waseda Institute for Advanced Study, Waseda University, Shinjuku, Tokyo 169-8050, Japan.}
\affiliation{Kavli Institute for Theoretical Sciences, University of Chinese Academy of Sciences, Beijing, 100190, China.}

\author{Hiroyuki Tajima}
\affiliation{Department of Physics, School of Science, The University of Tokyo, Tokyo 113-0033, Japan}

\author{Shun Uchino}
\affiliation{Faculty of Science and Engineering, Waseda University, Tokyo 169-8555, Japan}

\author{Mamoru Matsuo}
\affiliation{Kavli Institute for Theoretical Sciences, University of Chinese Academy of Sciences, Beijing, 100190, China.}
\affiliation{CAS Center for Excellence in Topological Quantum Computation, University of Chinese Academy of Sciences, Beijing 100190, China}
\affiliation{Advanced Science Research Center, Japan Atomic Energy Agency, Tokai, 319-1195, Japan}
\affiliation{RIKEN Center for Emergent Matter Science (CEMS), Wako, Saitama 351-0198, Japan}

\begin{abstract}
We investigate thermomagnetic transport in an ultracold atomic system with two ferromagnets linked via a magnetic quantum point contact.
Using nonequilibrium Green's function approach, we show a divergence in spin conductance and a slowing down of spin relaxation that manifest in the weak effective-Zeeman-field limit.
These anomalous spin dynamics result from the magnonic critical point at which magnons become gapless due to spontaneous magnetization.
Our findings unveil untapped dynamics in ultracold atomic systems, opening new avenues in thermomagnetism.
\end{abstract}

\maketitle

There is growing interest in the exploration of particle and heat flows in transport systems using ultracold atoms~\cite{sidorenkov2013second,brantut2013thermoelectric,krinner2017,husmann2018breakdown,PhysRevX.10.011042,christodoulou2021observation,hoffmann2021second,yan2022thermography,Krinner2016-gd}.
The ability to generate an ultraclean quantum many-body system in ultracold atom systems, free from impurity scattering or phonon effects typical in solid-state systems, establishes them as a prime stage for probing energy conversion mechanisms in instances where particle and heat flows intersect. This notion is reinforced by experiments at ETH~\cite{brantut2013thermoelectric,husmann2018breakdown} demonstrating that the thermoelectric response of a two-terminal system provides a controllable model system to navigate through the energy conversion mechanisms, thereby laying the groundwork for a ultracold-atom-based heat engine.

Recent investigations have seen a surge of interest in the field of spin transport using ultracold atoms~\cite{enss2019universal}, with studies revealing intriguing behaviors such as  spin diffusivity in a strongly-interacting Fermi gas~\cite{sommer2011universal,sommer2011spin} and super-exchange dynamics in Mott states realized with ultracold atoms in optical lattices~\cite{Nichols2019,Fukuhara2013-jv,Fukuhara2013-tj,Hild2014-hy,Jepsen2020-fg,Jepsen2021-ew,Jepsen2022-sf,Wei2022-qx}.
In light of the advanced methodologies now available for investigating spin transport, and in conjunction with the progress made in heat engine research mentioned earlier, there is an increasing impetus to study thermomagnetism, a realm where spin and heat flows intersect in two-terminal systems. In solid-state systems, these intersecting non-equilibrium phenomena form a unique field known as `spin caloritronics'~\cite{bauer2012spin}, where efficient spin-heat conversion mechanisms are actively examined from both theoretical and experimental perspectives~\cite{uchida2008observation,jaworski2010observation,uchida2010spin,xiao2010theory,adachi2011linear,adachi2013theory,ohnuma2017theory,matsuo2018spin,kato2019microscopic}. For instance, spin-heat transport phenomena at magnetic interfaces of junction systems comprised of magnetic materials and thin metal films are being energetically explored. However, these solid-state systems pose substantial challenges for the pursuit of efficient spin-heat conversion conditions due to complex factors such as interface disorder, various impurity scatterings, and relaxation processes mediated by phonons. 
In stark contrast, the ultracold atom systems provide an ideal platform for exploring spin-heat transport without such complex factors.
Indeed, bulk spin transport in ferromagnetic Heisenberg models has recently been explored in experiments of ultracold atoms~\cite{Fukuhara2013-jv,Fukuhara2013-tj,Hild2014-hy,Jepsen2020-fg,Jepsen2021-ew,Jepsen2022-sf,Wei2022-qx}.
\begin{figure}
    \centering
    \includegraphics[width=0.98\columnwidth,clip]{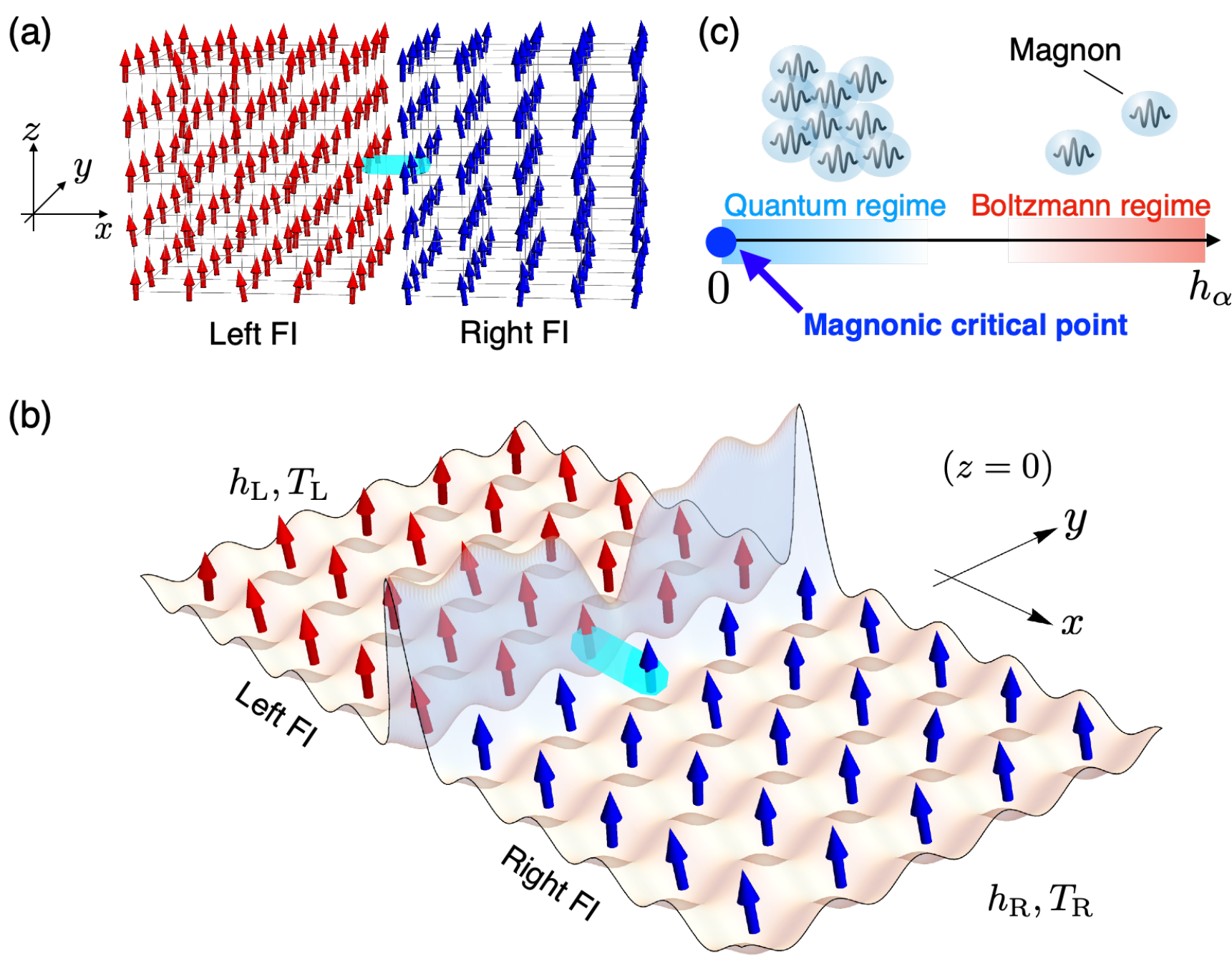}
    \caption{\label{fig:system}
    (a) Schematic for a magnetic quantum point contact (MQPC).
    Ferromagnetic insulators (FIs) in cubic lattices are connected by a weak exchange coupling represented by the light blue bond.
    (b) Potential profile in the plane $z=0$ to realize an MQPC in an optical lattice system. The barrier potential divides the system into left and right sections, allowing atoms to tunnel only through the light blue bond at the minimum of the barrier.
    (c) Phase diagram for magnons in FI at finite temperature with respect to an effective Zeeman field $h_\alpha$ along the $z$-axis.
    For small $h_\alpha$ (light blue region),     quantum statistics of magnons becomes crucial because of the overlap of their wave packets.
    The point $h_\alpha=0$ (blue blob) is the magnonic critical point at which magnons are gapless.
    }
\end{figure}

In this Letter, we propose that cold atoms are also suitable for investigating tunneling transport between ferromagnetic insulators (FIs) as shown in Fig.~\ref{fig:system} and microscopically explore the spin-heat cross-correlated transport. 
Using the Schwinger-Keldysh formalism, we develop a microscopic theory of spin and heat currents in a magnetic quantum point contact (MQPC). We reveal magnonic criticalities resulting in a divergently large spin conductance and enhanced cross-responses of heat and spin and a thermal conductance.
We also examine the relaxation dynamics of spin and heat to extract these properties in ultracold-atom experiments and find the slowing down of the spin relaxation.
Furthermore, we show that such divergence in spin conductance induces a non-linear spin current bias relation.
Our findings expand the scope of thermomagnetic transport research traditionally explored in solid-state physics, spintronics, and mesoscopic physics, leveraging the pristine and highly quantum-controllable nature of ultracold atomic systems.

{\it Model.---}
Hereafter we set $\hbar=k_\mathrm{B}=1$.
Let us consider magnon tunneling transport in the two-terminal system of FIs which can be realized with the Mott state of two-component bosonic atoms in an optical lattice~\cite{PhysRevLett.90.100401,Duan2003Controlling,Garcia-Ripoll:2003aa,Altman:2003aa,De_Hond2022-xi} depicted in Fig.~\ref{fig:system}.
The total Hamiltonian is given by
\begin{align}\label{eq:total_H}
\hat{\H}&=\hat{H}_\mathrm{L}+\hat{H}_\mathrm{R}+\hat{H}_\rmT,\\
\hat{H}_{\alpha=\rmL,\rmR}&=-J\sum_{\<\r_i,\r_j\>}{\hat{\bm{s}}}_{\r_i\alpha}\cdot{\hat{\bm{s}}}_{\r_j\alpha}-h_\alpha \hat{M}_\alpha,\\
\label{eq:H_T}
\hat{H}_\rmT&=-J_\rmT\,{\hat{\bm{s}}}_{\R_\rmL,\rmL}\cdot{\hat{\bm{s}}}_{\R_\rmR,\rmR}.
\end{align}
The bulk terms $\hat{H}_\alpha$ with the label $\alpha=\mathrm{L},\mathrm{R}$ describe the left (L) and right (R) FIs, where $J>0$ is the exchange coupling between nearest neighbors, ${\hat{\bm{s}}}_{\r_i\alpha}$ is a spin operator at the site $\r_i$ on a cubic lattice, $h_\alpha\geq0$ is the effective Zeeman field along the $z$ direction, and 
$\hat{M}_\alpha=\sum_{\r_i}\hat{s}^z_{\r_i\alpha}$ is the magnetization operator.
For simplicity, both FIs are assumed to have the same coupling constant $J$ and number of lattice sites $\N$.
We note that $h_\alpha$ are not external fields but effective ones controlled by the population imbalance between the two internal states of atoms~\cite{Krinner2016-gd,supplemental}.
The tunneling term $\hat{H}_\rmT$ describes the connection between the left and right FIs by a weak exchange coupling $J_\rmT\ll J$ denoted by the light blue bond in Fig.~\ref{fig:system}, where the sites $\R_\mathrm{L}$ and $\R_\mathrm{R}$ are edges of the bond.
Due to the barrier potential in Fig.~\ref{fig:system}(b), the system is partitioned into left and right regions, with atoms able to tunnel between them only via the light blue bond at the minimum of the barrier potential.
Because the deep Mott state is considered to realize FIs, the hopping of atoms over the bond results in the exchange coupling in Eq.~\eqref{eq:H_T}.
By analogy with a quantum point contact, which is a narrow construction for electronic or particle flows~\cite{krinner2017,nazarov}, we refer to this junction as a magnetic quantum point contact (MQPC).
In the case of highly polarized FIs as in Fig.~\ref{fig:system}, the MQPC is regarded as a quantum point contact for magnons~\cite{supplemental}.

We now investigate steady spin and heat currents, $I_\mathrm{S}$ and $I_\mathrm{H}$, in the case where both reservoirs are in thermal equilibrium at temperatures $T_\alpha$.
Similarly to tunneling transport of electrons or atoms~\cite{mahan2000many}, operators representing spin and heat currents are defined by $\hat{I}_\mathrm{S}(t)\equiv\d_t \hat{M}_\rmL$ and $\hat{I}_\mathrm{H}(t)\equiv-\d_t \hat{H}_\mathrm{L}$, respectively. 
We evaluate $I_\mathrm{S}$ and $I_\mathrm{H}$ by taking the thermal average of $\hat{I}_\mathrm{S}(t)$ and $\hat{I}_\mathrm{H}(t)$ and performing the perturbative expansion in $\hat{H}_\rmT$ up to $\order{(J_\mathrm{T})^2}$ in the Schwinger-Keldysh formalism~\cite{ohnuma2017theory,kato2019microscopic,nakata2018magnonic,Nakata2015-ns}.
Because we are considering the situation where both FIs are highly magnetized in the $+z$ direction as in Figs.~\ref{fig:system} (a) and (b), we can employ the spin-wave approximation to compute correlation functions, which are relevant to the currents in the case of $0<T_\alpha\ll T_\mathrm{c}$ with $T_\mathrm{c}\sim J$ being the Curie temperature~\cite{holstein1940field,Dyson1956-db}.
As a result, the currents can be expressed in terms of magnons, which are quasiparticles governing low-energy excitations in highly polarized FIs.
Indeed, $I_\mathrm{S}$ and $I_\mathrm{H}$ are found to be~\cite{supplemental}
\begin{align}
\label{eq:I_S_2}
I_\mathrm{S}
&=\int_{-\infty}^\infty\!\!d\omega\,\T(\omega)\varDelta n_\mathrm{B}(\omega),\\
\label{eq:I_H_2}
I_\mathrm{H}
&=\int_{-\infty}^\infty\!\!d\omega\,(\omega+h_\mathrm{L})\,\T(\omega)\varDelta n_\mathrm{B}(\omega),
\end{align}
respectively, where 
$\T(\omega)=\sum_{\p\q}\frac{(J_\mathrm{T}/\N)^2}{2\pi}\A_{\p\mathrm{L}}(\omega+h_\mathrm{L})\A_{\q\mathrm{R}}(\omega+h_\mathrm{R})$ 
is the energy-dependent magnon transmittance,
$\A_{\k\alpha}(\omega)=\pi\delta(\omega-E_{\k\alpha})$ 
is the spectral function of a magnon with energy $E_{\k\alpha}\simeq h_\alpha+(J/2)\k^2$ and dimensionless momentum $\k$, and
$\varDelta n_\mathrm{B}(\omega)=n_{\mathrm{B,L}}(\omega+h_\mathrm{L})-n_{\mathrm{B,R}}(\omega+h_\mathrm{R})$
denotes the difference in the Bose-Einstein (BE) distribution function of magnons $n_{\mathrm{B},\alpha}(\omega)=1/(e^{\omega/T_\alpha}-1)$ between left and right FIs.
Equations~\eqref{eq:I_S_2} and \eqref{eq:I_H_2} mean that spin and heat are transported by magnons tunneling between FIs.
In our case of an MQPC, 
$I_\mathrm{S}$ and $I_\mathrm{H}$ can be analytically computed.
After straightforward evaluations of Eqs.~\eqref{eq:I_S_2} and \eqref{eq:I_H_2}, we obtain
\begin{align}\label{eq:I_S_SWA}
I_\mathrm{S}
&=A\varDelta F_{2},\\
\label{eq:I_H_SWA}
I_\mathrm{H}
&=A\qty(2\varDelta F_{3}+h_\mathrm{L}\varDelta F_{2}),
\end{align}
where $A=\frac{(J_\mathrm{T})^2}{4\pi^{3}J^3}$,
$\varDelta F_d=(T_\rmL)^d F_{d}(x_\rmL)-(T_\rmR)^d F_{d}(x_\rmR)$,
$F_{d}(x_\alpha)= \mathrm{Li}_d\qty(e^{-x_\alpha})$
is the BE integral with $x_\alpha=h_\alpha/T_\alpha$~\cite{supplemental,Robinson1951note,takahashi1986quantum}, and $\mathrm{Li}_d(z)$ is the polylogarithm.
The effects of the magnon distributions on $I_\mathrm{S}$ and $I_\mathrm{H}$ are encoded in 
$F_d(x_\alpha)\propto\int_0^\infty\!\!\dd{\omega}\omega^{d-1}n_{\mathrm{B},\alpha}(\omega+h_\alpha)$.
Because of the divergence of $n_{\mathrm{B},\alpha}(\omega)\sim1/\omega$ at $\omega=0$, $F_{d}(x_\alpha)$ 
behaves nonanalytically near 
$x_\alpha=0$.
As discussed below, this singular behavior results from the criticality of magnons at $x_\alpha=0$.

{\it Quantum regime and magnonic criticality.---}
Here, we discuss the quantum features of magnons including the magnonic criticality.
As shown in  Fig.~\ref{fig:system}(c), the quantum nature of magnons becomes more pronounced as $h_\alpha$ decreases with $T_\alpha$ fixed.
This can be intuitively understood by comparing $T_\alpha$ with the energy gap of magnons $E_{\mathrm{gap},\alpha}=h_\alpha$ given by $h_\alpha$.
In the small-field regime  $h_\alpha\ll T_\alpha$, where the temperature is larger than the gap, so many magnons are thermally excited that their wave packets overlap with each other.
Therefore, the BE statistics of magnons becomes essential.
For this reason, the regime $h_\alpha\ll T_\alpha$ is referred to as the quantum regime in this Letter.
For $h_\alpha\gg T_\alpha$, on the other hand, $E_{\mathrm{gap},\alpha}\gg T_\alpha$ suppresses thermal excitation of magnons, which behave as a dilute Boltzmann gas.

The zero-field case $h_\alpha=0$ depicted as a blue blob in Fig.~\ref{fig:system}(c) corresponds to the critical point for magnons with the vanishing energy gap.
In this case, FI is spontaneously magnetized in the absence of $h_\alpha$.
The spontaneous symmetry breaking of spin rotations makes magnons behave as the gapless Nambu-Goldstone mode~\cite{watanabe2012unified,hidaka2013counting}.
In this way, the magnonic criticality at $h_\alpha=0$ is related to closing of the magnon gap due to the spontaneous magnetization.

Due to the magnonic criticality,  transport and thermodynamic quantities of FIs exhibits singular behaviors in the quantum regime $h_\alpha\ll T_\alpha$.
Near the critical point $x_\alpha=h_\alpha/T_\alpha\to+0$, $F_d(x_\alpha)$ 
behaves nonanalytically, leading to singularities of the currents in Eqs.~\eqref{eq:I_S_SWA} and \eqref{eq:I_H_SWA}. 
In addition, the BE integrals also characterize critical behaviors of thermodynamic quantities in FI~\cite{supplemental}.
For example, the differential spin susceptibility $\kappa_\alpha=\qty(\pdv{M_\alpha}{h_\alpha})_{T_\alpha}\propto F_{1/2}(x_\alpha)$ with $M_\alpha$ being the magnetization shows a divergent behavior $\kappa_\alpha\sim1/\sqrt{h_\alpha}$ in the limit $h_\alpha\to+0$.
The same form of divergence also appears in the isothermal compressibility of an ideal Bose gas above 
the BE condensation temperature~\cite{pitaevskii2003}.
In this way, $F_d(x_\alpha)$ describe the critical behaviors of physical quantities in bosonic systems.
Below, we discuss in detail how the criticality associated with the gapless magnons
affects
the spin and heat currents in Eqs.~\eqref{eq:I_S_SWA} and \eqref{eq:I_H_SWA}.

{\it Onsager coefficients.---}
First, we investigate the quantum effects on transport in a small-bias regime.
By expanding Eqs.~\eqref{eq:I_S_SWA} and \eqref{eq:I_H_SWA} in spin and temperature biases, $\varDelta h=-(h_\rmL-h_\rmR)$ and $\varDelta T=T_\rmL-T_\rmR$, and using $\dv{y}F_d(y)=-F_{d-1}(y)$, $I_\mathrm{S}$ and $I_\mathrm{H}$ are expressed as
\begin{align}\label{eq:linear-response}
\begin{pmatrix}
I_\mathrm{S}\\
I_\mathrm{H}
\end{pmatrix}
=
\begin{pmatrix}
L_{11}&L_{12}\\
L_{21}&L_{22}
\end{pmatrix}
\begin{pmatrix}
\varDelta h\\
\varDelta T
\end{pmatrix},
\end{align}
where Onsager coefficients are given by
\begin{align}\label{eq:L_11}
\frac{L_{11}}{AT}
&=F_1(x),\\
\label{eq:L_12}
\frac{L_{12}}{AT}
&=\frac{L_{21}}{AT^2}
=2\,F_{2}(x)+x\,F_{1}(x),\\
\label{eq:L_22}
\frac{L_{22}}{AT^{2}}
&=6\,F_{3}(x)+4x\,F_{2}(x)+x^2\,F_{1}(x),
\end{align}
$x=h/T$, $h=(h_\rmL+h_\rmR)/2$ and $T=(T_\rmL+T_\rmR)/2$ are the averaged effective Zeeman field and temperature, respectively.
As in Eq.~\eqref{eq:L_12}, the cross-response terms satisfy the reciprocal relation $L_{21}=L_{12}T$.

\begin{figure}
    \centering
    \includegraphics[width=0.98\columnwidth,clip]{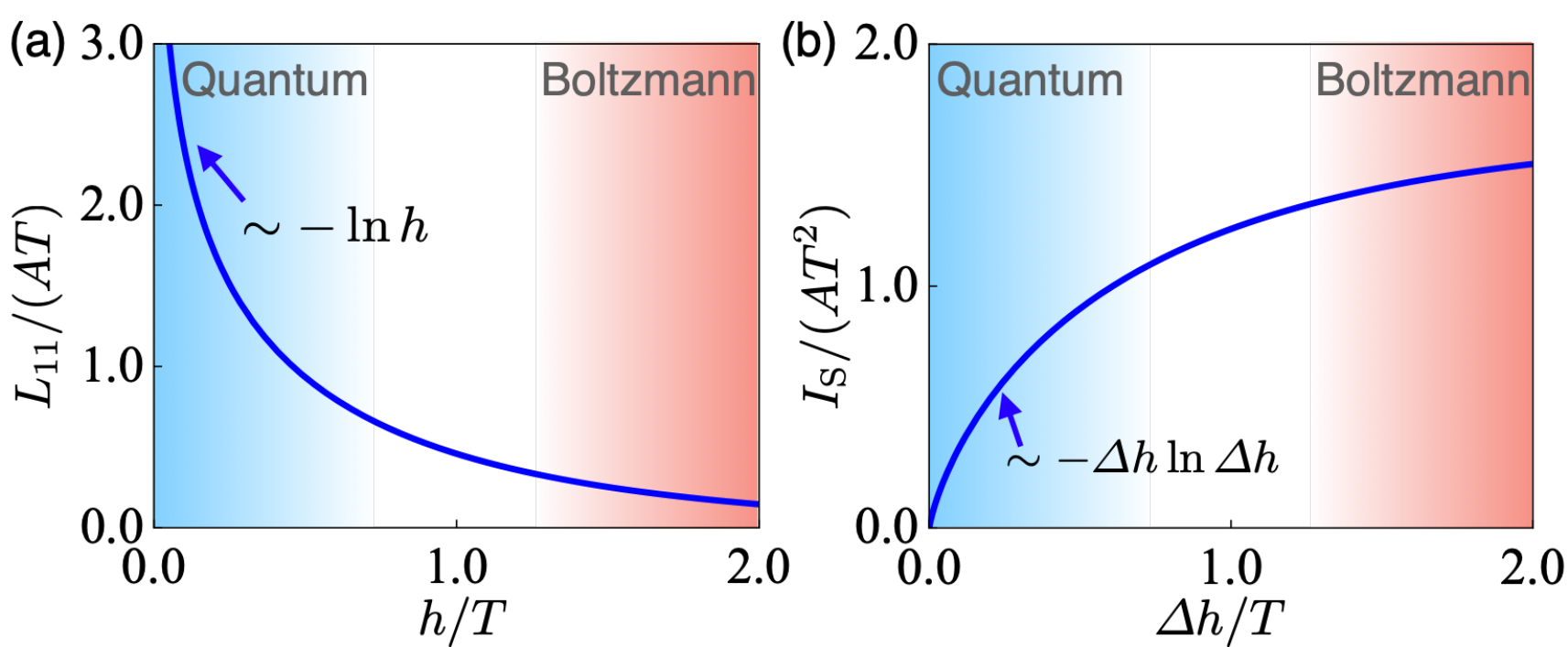}    
    \caption{Quantum spin transport resulting from the magnonic criticality at $h_{\rmL,\rmR}=0$.
    (a) Spin conductance $L_{11}$ as a function of the mean effective Zeeman field $h=(h_\rmL+h_\rmR)/2$.
    In the quantum regime (blue region), $L_{11}$ exhibits divergent behaviors, which implies breakdown of Ohm's law for spin.
    (b) Nonlinear current-bias characteristic of spin at $h_\rmL=\varDelta T=0$.
    In the quantum regime, the spin current $I_\mathrm{S}$ nonlinearly depends on the spin bias $\varDelta h=h_\rmR$ because of the criticality.
    }
    \label{fig:transport}
\end{figure}
The spin conductance $L_{11}$ is sensitive to the magnonic criticality and shows a divergent behavior in the quantum regime ($x=h/T\ll1$).
As shown in Fig.~\ref{fig:transport}(a), $L_{11}$ increases as $h$ becomes smaller with $T$ fixed.
This can be regarded as the enhancement of a spin current due to quantum degeneracy.
Because of $E_{\mathrm{gap},\alpha}\simeq h$ as discussed above, the number of thermally excited magnons [see Fig.~\ref{fig:system}(c)] contributing to a spin current increases with decreasing $h$, so that $L_{11}$ is enhanced.
In addition, Fig.~\ref{fig:transport}(a) represents a divergent behavior of $L_{11}$ near the critical point $h=0$, at which the magnon gap is closed due to spontaneous magnetization.
Indeed, the magnonic criticality leads to the logarithmic divergence of the spin conductance in Eq.~\eqref{eq:L_11}:
\begin{align}\label{eq:L_11_critical}
\frac{L_{11}}{AT}&=-\ln \left(\frac{h}{T}\right).
\end{align}
The divergence of $L_{11}$ implies that Ohm's law $I_\mathrm{S}=L_{11}\varDelta h$ is broken when both FIs are near the critical points as discussed later.

Quantum degeneracy of magnons enhances not only the spin conductance $L_{11}$ but also the cross-responses of spin and heat, $L_{12}$ and $L_{21}$, and diagonal heat response $L_{22}$ in the quantum regime.
Similarly to $L_{11}$ in Fig.~\ref{fig:transport}(a), $L_{12}$ and $L_{21}$ in Eq.~\eqref{eq:L_12} and $L_{22}$ in Eq.~\eqref{eq:L_22} monotonically increase as $h\geq0$ decreases with $T$ fixed~\cite{supplemental}.
At the magnonic critical point $h=0$, however, $L_{12}$, $L_{21}$, and $L_{22}$ are convergent, while $L_{11}$ is divergent as in Eq.~\eqref{eq:L_11_critical}.
Indeed, $L_{12}/(AT)=L_{21}/(AT^2)=\pi^2/3$ and $L_{22}/(AT^{2})=6\zeta(3)$ are obtained at $h=0$.

{\it Breakdown of the Ohm's law for magnons.---}
To further understand the impact of the criticality on magnonic transport, we focus on the case where the left FI is at the critical point $h_\rmL=0$ with spontaneous magnetization.

In this case, a spin current driven by the spin bias violates Ohm's law.
Figure~\ref{fig:transport}(b) shows that $I_\mathrm{S}$ nonlinearly depends on $\varDelta h=h_\rmR$ in the isothermal condition $\varDelta T=0$.
When the right FI is in the quantum regime $\varDelta h/T\ll1$, $I_\mathrm{S}$ exhibits nonanalytic behaviors as
\begin{align}\label{eq:I_S_h=0}
\frac{I_\mathrm{S}}{AT^{2}}=-\frac{\varDelta h}{T}\ln \left(\frac{\varDelta h}{T}\right).
\end{align}
Such a logarithmic correction arises from the BE integral $F_{2}(\varDelta h/T)$ near the magnonic critical point, leading to the logarithmic divergence of $L_{11}$ in Eq.~\eqref{eq:L_11_critical}.
We note that this nonlinear characteristic resulting from the criticality is specific to the spin current driven by the spin bias.
Indeed, thermally-driven spin current is linear in $\varDelta T$ for a small temperature bias even when both FIs are at the critical points $h_\rmL=h_\rmR=0$.
This corresponds to the convergence of $L_{12}$ at $h=0$.
Similarly, the convergences of $L_{21}$ and $L_{22}$ at $h=0$ ensure that the heat current is linear in both spin and heat biases.

{\it Relaxation dynamics.---}
\begin{figure*}[tb]
    \centering
    \includegraphics[width=2\columnwidth,clip]{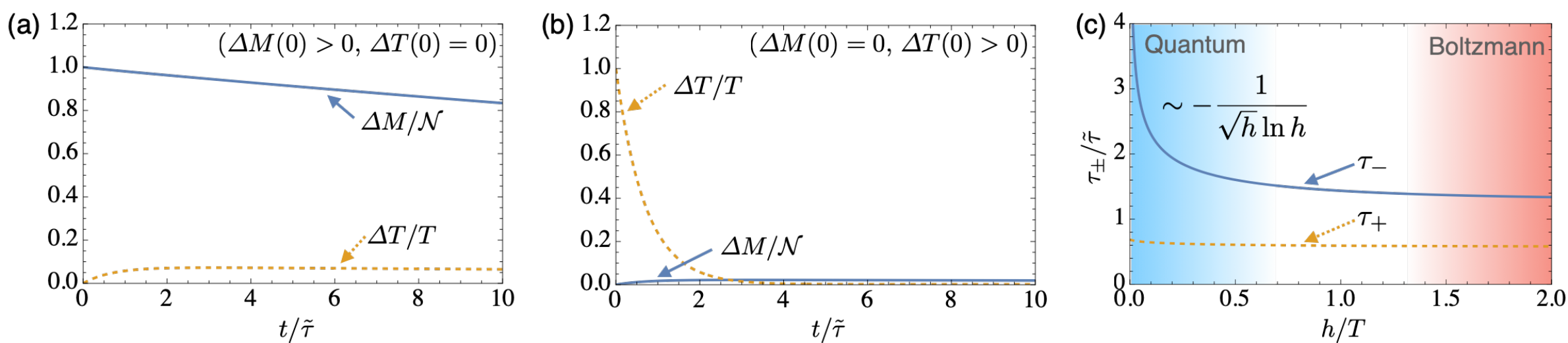}
    \caption{\label{fig:relaxation}
    Relaxation dynamics of spin and heat in the vicinity of the critical point.
    (a) [(b)] Dynamics of the magnetization imbalance $\varDelta M(t)$ and temperature bias $\varDelta T(t)$ between FIs driven by $\varDelta M(0)>0$  [$\varDelta T(0)>0$] at initial time 
    with $T/J=0.30$ and $\sqrt{hJ^3/T^4}=0.19$.
    Because of the criticality, $\varDelta M(t)$ in (a) shows extremely slow relaxation with the decay time $\tau_-=56\tilde{\tau}$ compared with $\varDelta T(t)$ in (b) with the decay time $\tau_+=0.70\tilde{\tau}$.
    In addition, the criticality results in suppression of the cross-responses, $\varDelta T(t)$ in (a) and $\varDelta M(t)$ in (b).
    The vertical axes in (a) and (b) are normalized by $\varDelta M(0)/\N$ and $\varDelta T(0)/T$, respectively.    
    (c) Decay times $\tau_\pm$ as functions of $h$. In the limit $h\to+0$, $\tau_-$ becomes divergent due to the magnonic criticality, while $\tau_+\to0.70\tilde{\tau}$.
    }
\end{figure*}
Here, we discuss relaxation dynamics of spin and heat which allows one to extract transport properties $L_{ij}$ in ultracold-atom experiments.

Since ultracold atoms are isolated, initially prepared biases gradually vanish by exchanging spin and heat between reservoirs.
This situation is different from that in solid-state systems where biases can be externally kept constant and thus Eq.~\eqref{eq:linear-response} is directly applicable.
The scheme to experimentally extract $L_{ij}$ in isolated two-terminal systems has been established in the context of ultracold fermionic atoms~\cite{brantut2013thermoelectric,grenier2012probing}.
Here, we apply this method to our system of FIs with an MQPC.

In this scheme, thermalized FIs with given initial magnetizations and temperatures are first prepared and they are completely decoupled by closing the channel of the MQPC with the enhanced barrier potential.
At time $t=0$, the channel is opened and spin and heat are exchanged between FIs by the tunneling term in Eq.~\eqref{eq:H_T}.
Then, the relaxation of the magnetization imbalance $\varDelta M(t)=M_\rmL(t)-M_\rmR(t)$ and temperatures bias $\varDelta T(t)=T_\rmL(t)-T_\rmR(t)$ is observed.
We assume that the initial values $|\varDelta M(0)|\ll\N$ and $|\varDelta T(0)|\ll T=[T_\rmL(0)+T_\rmR(0)]/2$ are small and that the relaxation time scales of $\varDelta M(t)$ and $\varDelta T(t)$ are much longer than the thermalization time $\tau_\mathrm{th}$ within each FI.
In such a situation, the relaxation can be analyzed under the quasi-stationary condition where FIs at each $t>0$ are regarded as thermalized states with $M_\alpha(t)$ and $T_\alpha(t)$, leading to simple differential equations governing time evolutions of $\varDelta M(t)$ and $\varDelta T(t)$.
As a result, the solutions of $\varDelta M(t)$ and $\varDelta T(t)$ are formally written in terms of $L_{ij}$ and equilibrium thermodynamic quantities such as the differential spin susceptibility $\kappa=\qty(\pdv{M}{h})_T$
with $M=[M_\rmL(0)+M_\rmR(0)]/2$ and $h=[h_\rmL(0)+h_\rmR(0)]/2$~\cite{brantut2013thermoelectric,grenier2012probing,supplemental}.
In addition, the thermodynamic quantities in equilibrium are experimentally accessible independently from the relaxation dynamics~\cite{Takasu2020Energy}.
Therefore, by observing both relaxation dynamics and equilibrium thermodynamics and using the formal solutions of $\varDelta M(t)$ and $\varDelta T(t)$, one can experimentally extract $L_{ij}$.

To demonstrate the relaxation dynamics, we evaluate the time evolution of $\varDelta M(t)$ and $\varDelta T(t)$ with the spin-wave approximation and find that the magnonic criticality results in extremely slow spin relaxation~\cite{supplemental}.
Figure~\ref{fig:relaxation}(a) [(b)] shows dynamics of $\varDelta M(t)$ and $\varDelta T(t)$ driven by an initial magnetization imbalance (temperature bias) near the magnonic critical point at low temperature.
The parameters $h$ and $T$ are set to the same values in both panels.
In Fig.~\ref{fig:relaxation}(a), the magnetization imbalance $\varDelta M(t)\simeq e^{-t/\tau_-}\varDelta M(0)$ relaxes very slowly with a decay time $\tau_-=56\tilde{\tau}$ much larger than the typical relaxation time scale $\tilde{\tau}=\sqrt{\pi^3/2}\N(J/J_\mathrm{T})^2/\sqrt{JT}$~\footnote{Because of $\tilde{\tau}\gg1/J\sim\tau_\mathrm{th}$ in our case with $\N,\, J/J_\rmT,\, J/T\gg1$, the quasi-stationary treatment is valid.}.
Indeed, this spin relaxation is much slower than the temperature-bias relaxation $\varDelta T(t)\simeq e^{-t/\tau_+}\varDelta T(0)$ in Fig.~\ref{fig:relaxation}(b) with a decay time $\tau_+=0.70\tilde{\tau}$.
To see the impact of the magnonic criticality on the dynamics in details, we plot $\tau_\pm$ as functions of $h/T$ in Fig.~\ref{fig:relaxation}(c).
For $h\to+0$, $\tau_-$ exhibits a divergence $\tau_-\sim \tilde{\tau}/[\sqrt{h}\ln (h/T)]$, while $\tau_+\to0.70\tilde{\tau}$ converges.
The divergence of $\tau_-\sim \kappa/L_{11}$ near the magnonic critical point arises from the fact that $\kappa\propto F_{1/2}(x)\sim1/\sqrt{h}$ diverges faster than $L_{11}\sim-\ln (h/T)$ in Eq.~\eqref{eq:L_11_critical}.
In this way, the critical behavior of the differential spin susceptibility results in slowing down of the spin relaxation.

In addition, the magnonic criticality makes time evolution of $\varDelta M(t)$ and $\varDelta T(t)$ almost uncorrelated.
In the limit of $h/T\to0$ at low temperature $0<T\ll J$, we find that $\varDelta T(t)\sim\sqrt{h/T}$ induced by the initial magnetization imbalance in Fig.~\ref{fig:relaxation}(a) and $\varDelta M(t)\sim(T/J)^{3/2}$ induced by the initial temperature bias in Fig.~\ref{fig:relaxation}(b) are negligibly small.
Therefore, the magnonic criticality diminishes the cross-responses of $\varDelta M(t)$ and $\varDelta T(t)$.
We stress that this dynamics resulting from the magnonic criticality is remarkable since $\varDelta M(t)$ and $\varDelta T(t)$
are normally correlated with each other~\cite{brantut2013thermoelectric}.

It is worth mentioning the following points.
First, dynamics with $\tau_-\gg\tau_+$ has also been observed in a two-terminal system of unitary Fermi gases whose microscopic description has yet to be resolved~\cite{husmann2018breakdown}.
Second, details of the junction geometry affect the relaxation time as well as transport properties in bosonic two-terminal systems near a critical point~\cite{Papoular2016quantized,sekino_transport_inprep}.
Third, similar transport and relaxation dynamics may be achieved with weakly interacting Bose gases near the BE critical temperature with a quantum point contact~\cite{supplemental,sekino_transport_inprep}.
Finally, $h_\alpha$ at $t<0$ can be experimentally controlled by tuning population imbalance of bosonic atoms~\cite{supplemental,Takasu2020Energy}.

{\it Conclusion.---}
In this Letter, we discussed spin and heat transports in two-terminal systems consisting of two FIs connected via a MQPC, which can be realized with ultracold bosonic atoms in an optical lattice at accessible temperature scale~\cite{Chiu2019-up,Shao2024-uc}.
We found the anomalous impacts of the magnonic criticality at the zero effective Zeeman field, resulting in the breakdown of Ohm's law for spin and the extremely slowing down of spin relaxation. 
Our findings uncover untapped potentials in ultracold atomic systems, paving the way for new advancements in 
thermomagnetism.

\begin{acknowledgments}
The authors are grateful to T. Fukuhara for useful discussions on experimental realization and T. Kato for support in the early stage of this work.
 We acknowledge JSPS KAKENHI for Grants (JP20H01863, JP21H04565,  JP21H01800, JP21K03436,
 JP22H01158,
 JP22K13981). 
 Y. S. is supported by Pioneering Program of RIKEN for Evolution of Matter in the Universe
(r-EMU) and by JST ERATO Grant Number JPMJER2302, Japan.
S. U. is supported  by JST PRESTO (JPMJPR2351),
MEXT Leading Initiative for Excellent Young Researchers (JPMXS032020000), and Matsuo Foundation.
M. M. is supported by the Priority Program of the Chinese Academy of Sciences, Grant No. XDB28000000.
\end{acknowledgments}

\bibliography{masterbib}

\widetext
\newpage
\newpage

\begin{center}
\textbf{\large Supplemental Materials: Thermomagnetic anomalies by magnonic criticality in ultracold atomic transport}
\end{center}
 \setcounter{equation}{0}
 \setcounter{figure}{0}
 \setcounter{table}{0}
 \setcounter{page}{1}
\makeatletter
 \renewcommand{\theequation}{S\arabic{equation}}
 \renewcommand{\thefigure}{S\arabic{figure}}
 \renewcommand{\bibnumfmt}[1]{[S#1]}

\section{Tunneling currents}
Here, we analyze tunneling currents.
For convenience, we define the Hamiltonian $\hat{\H}^{(0)}=\hat{H}^{(0)}_\mathrm{L}+\hat{H}^{(0)}_\mathrm{R}+\hat{H}_\rmT$ with $\hat{H}^{(0)}_\alpha\equiv-J\sum_{\<\r_i,\r_j\>}{\hat{\bm{s}}}_{\r_i\alpha}\cdot{\hat{\bm{s}}}_{\r_j\alpha},$ in the absence of the Zeeman term.
Spin, energy, and heat currents flowing from left to right FIs are described by the operators
\begin{align}
\hat{I}_\mathrm{S}(t)&\equiv\d_t \hat{M}_\mathrm{L}(t)=-i[\hat{M}_\mathrm{L}(t),\hat{\H}^{(0)}(t)]=-i[\hat{M}_\mathrm{L}(t),\hat{H}_\rmT(t)],\\
\hat{I}_\mathrm{E}(t)&\equiv-\d_t \hat{H}^{(0)}_\mathrm{L}(t)=i[\hat{H}^{(0)}_\mathrm{L}(t),\hat{\H}^{(0)}(t)]=i[\hat{H}^{(0)}_\mathrm{L}(t),\hat{H}_\rmT(t)],\\
\hat{I}_\mathrm{H}(t)&\equiv-\d_t \hat{H}_\mathrm{L}(t)
=\hat{I}_\mathrm{E}(t)+h_\mathrm{L}\hat{I}_\mathrm{S}(t),
\end{align}
where $\hat{O}(t)$ is the Heisenberg representation of the operator $\hat{O}$ with respect to $\hat{\H}^{(0)}$.
We note that this situation is similar to the tunneling Hamiltonian approach for a two-terminal electron system, where time evolution of a physical quantity is governed by the Hamiltonian without chemical potential terms~\cite{mahan2000many}.
To evaluate tunneling currents, we use the commutation relations for spin operators and take the thermal average with respect to the density operator
\begin{align}\label{eq_sup:rho_H}
\hat{\rho}&=\hat{\rho}_{\hat{H}_\mathrm{L}}\otimes\hat{\rho}_{\hat{H}_\mathrm{R}}=\frac{e^{-\hat{H}_\rmL/T_\rmL}}{\Tr[e^{-\hat{H}_\rmL/T_\rmL}]}\otimes\frac{e^{-\hat{H}_\rmR/T_\rmR}}{\Tr[e^{-\hat{H}_\rmR/T_\rmR}]}.
\end{align}
Note that $\hat{\rho}$ includes the effective Zeeman fields $h_\alpha$ via $\hat{H}_\alpha=\hat{H}^{(0)}_\alpha-h_\alpha\hat{M}_\alpha$ as that for two-terminal electron systems includes chemical potentials~\cite{mahan2000many}.
The currents are then obtained as
\begin{align}
I_\mathrm{S}(t)&=\<\hat{I}_\mathrm{S}(t)\>
=\Im\,[\I_{+-}^<(t_1,t_2)]_{t_1,t_2\to t},\\
I_\mathrm{E}(t)&=\<\hat{I}_\mathrm{E}(t)\>
=-\Re[\d_{t_2}\I_{+-}^<(t_1,t_2)]_{t_1,t_2\to t}-[\d_{t_2}\I_{zz}^<(t_1,t_2)]_{t_1,t_2\to t},\\
I_\mathrm{H}(t)&=\<\hat{I}_\mathrm{H}(t)\>
=I_\mathrm{E}(t)+h_\mathrm{L}I_\mathrm{S}(t),
\end{align}
where 
\begin{align}
    \I_{+-}^<(t_1,t_2)&=-\sum_{\p\q}\mathcal{J}_{\p\q}\<\hat{s}_{-\p\mathrm{L}}^-(t_2)\hat{s}_{\q\mathrm{R}}^+(t_1)\>,\\
    \I_{zz}^<(t_1,t_2)&=-\sum_{\p\q}\mathcal{J}_{\p\q}\<\hat{s}_{-\p\mathrm{L}}^z(t_2)\hat{s}_{\q\mathrm{R}}^z(t_1)\>,
\end{align}
$\mathcal{J}_{\p\q}=\frac{-J_\mathrm{T}}{\N}e^{-i\p\cdot\R_\rmL+i\q\cdot\R_\rmR}$, 
${\hat{\bm{s}}}_{\r_i\alpha}=\frac{1}{\sqrt{\N}}\sum_\p e^{i\p\cdot\r_i}\hat{\bm{s}}_{\p\alpha}$, 
$\hat{s}_{\k\alpha}^\pm=\hat{s}_{\k\alpha}^x\pm i\hat{s}_{\k\alpha}^y$, and $\<\cdots\>=\Tr\left[\hat{\rho}\cdots\right]$.
We perturbatively calculate the currents up to the second order of $J_\rmT$.
Such computation can be performed by employing the Schwinger-Keldysh formalism, where $\I_{+-}^<(t_1,t_2)$ and $\I_{zz}^<(t_1,t_2)$ are regarded as lesser components of contour correlation functions~\cite{ohnuma2017theory,kato2019microscopic,nakata2018magnonic,Nakata2015-ns}.
After the perturbative calculations, we obtain
\begin{align}\label{eq_sup:I_S_1}
I_\mathrm{S}
&=\int_{-\infty}^\infty\!\!\dd{\omega}\T_{+-}(\omega)\varDelta n_\mathrm{B}(\omega),\\
\label{eq_sup:I_E_1}
I_\mathrm{E}
&=\int_{-\infty}^\infty\!\!\dd{\omega}\omega\,\T_{+-}(\omega)\varDelta n_\mathrm{B}(\omega)
+\int_{-\infty}^\infty\!\!\dd{\omega}\omega\,\T_{zz}(\omega)\varDelta n_\mathrm{B}^{(0)}(\omega),\\
\label{eq_sup:I_H_1}
I_\mathrm{H}
&=I_\mathrm{E}+h_\mathrm{L}I_\mathrm{S}.
\end{align}
The transmittance functions $\T_{+-}$ and $\T_{zz}$ are given by
\begin{align}\label{eq_sup:G_+-(omega)}
\T_{+-}(\omega)&=\sum_{\p\q}\frac{|\mathcal{J}_{\p\q}|^2}{2\pi}\Im\,G_{\p\mathrm{L}}^{+-;\mathrm{ret}}(\omega+h_\mathrm{L})\Im\,G_{\q\mathrm{R}}^{+-;\mathrm{ret}}(\omega+h_\mathrm{R}),\\
\label{eq_sup:G_zz(omega)}
\T_{zz}(\omega)&=\sum_{\p\q}\frac{|\mathcal{J}_{\p\q}|^2}{\pi}\Im\,G_{\p\mathrm{L}}^{zz;\mathrm{ret}}(\omega)\Im\,G_{\q\mathrm{R}}^{zz;\mathrm{ret}}(\omega),
\end{align}
respectively, in terms of the retarded response functions of spins
\begin{align}\label{eq_sup:G^+-_ret}
G_{\k\alpha}^{+-;\mathrm{ret}}(\omega)&=-i\int_0^\infty\!\!\dd{t}e^{i\omega^+t}\<[\hat{s}_{\k\alpha;\hat{H}_\alpha}^+(t),\hat{s}_{-\k\alpha;\hat{H}_\alpha}^-(0)]\>,\\
\label{eq_sup:G^zz_ret}
G_{\k\alpha}^{zz;\mathrm{ret}}(\omega)&=-i\int_0^\infty\!\!\dd{t}e^{i\omega^+t}\<[\hat{s}_{\k\alpha;\hat{H}_\alpha}^z(t),\hat{s}_{-\k\alpha;\hat{H}_\alpha}^z(0)]\>,
\end{align}
where $\omega^+=\omega+i0^+$ and 
$\hat{\bm{s}}_{\k\alpha;\hat{H}_\alpha}(t)=e^{i\hat{H}_\alpha t}\hat{\bm{s}}_{\k\alpha}e^{-i\hat{H}_\alpha t}$ is the Heisenberg representation with respect to $\hat{H}_\alpha$ in the presence of the effective Zeeman term.
In Eqs.~\eqref{eq_sup:I_S_1}--\eqref{eq_sup:I_H_1}, $\varDelta n_\mathrm{B}(\omega)=n_{\mathrm{B,L}}(\omega+h_\mathrm{L})-n_{\mathrm{B,R}}(\omega+h_\mathrm{R})$ and $\varDelta n_\mathrm{B}^{(0)}(\omega)=n_{\mathrm{B,L}}(\omega)-n_{\mathrm{B,R}}(\omega)$ are the difference of the Bose distribution function $n_{\mathrm{B},\alpha}(\omega)=1/(e^{\omega/T_\alpha}-1)$ between the two FIs.
The frequency shifts by the effective Zeeman fields appear in $\T_{+-}(\omega)$ and $\varDelta n_\mathrm{B}(\omega)$, while $\T_{zz}(\omega)$ and $\varDelta n_\mathrm{B}^{(0)}(\omega)$ associated with $\I_{zz}^<(t_1,t_2)$ do not include the shift.
The presence and absence of the shift result from $\hat{s}_{\k\alpha;\hat{H}^{(0)}_\alpha}^\pm(t)=\hat{s}_{\k\alpha;\hat{H}_\alpha}^\pm(t)e^{\pm ih_\alpha t}$ and $\hat{s}_{\k\alpha;\hat{H}^{(0)}_\alpha}^z(t)=\hat{s}_{\k\alpha;\hat{H}_\alpha}^z(t)$.
To go further, we employ the spin-wave approximation and evaluate the currents in Eqs.~\eqref{eq_sup:I_S_1}--\eqref{eq_sup:I_H_1}.

\section{Spin-wave approximation}
Here, we employ the spin-wave approximation, which is justified when the two FIs are highly spin polarized along the $+z$-direction at low temperature.
We start with the the Holstein-Primakoff transformations given by~\cite{holstein1940field}
\begin{align}\label{eq_sup:HP_S^-}
\hat{s}_{\r_i\alpha}^-&=\hat{b}_{\r_i\alpha}^\+(2\S-\hat{b}_{\r_i\alpha}^\+\hat{b}_{\r_i\alpha})^{1/2},\\
\label{eq_sup:HP_S^+}
\hat{s}_{\r_i\alpha}^+&=(2\S-\hat{b}_{\r_i\alpha}^\+\hat{b}_{\r_i\alpha})^{1/2}\hat{b}_{\r_i\alpha},\\
\label{eq_sup:HP_S^z}
\hat{s}_{\r_i\alpha}^z&=\S-\hat{b}_{\r_i\alpha}^\+\hat{b}_{\r_i\alpha},
\end{align}
where the c-number $\S=1/2,1,\cdots$ satisfying $({\bm \hat{s}}_{\r_i\alpha})^2=\S(\S+1)$ represents the magnitude of spin and $\hat{b}_{\r_i\alpha}$ and $\hat{b}_{\r_i\alpha}^\+$ are the annihilation and creation operators satisfying the commutation relations for bosons.
The degrees of freedom of bosons describes the spin fluctuations.
It is convenient to introduce $\S$ because the the spin-wave approximation is formally associated with the expansion with respect to $1/\S$.
Expanding Eqs.~\eqref{eq_sup:HP_S^-}--\eqref{eq_sup:HP_S^z} in $1/\S$ yields
\begin{align}\label{eq_sup:HP_S^-_SWA}
\hat{s}_{\r_i\alpha}^-&=\sqrt{2\S}\hat{b}_{\r_i\alpha}^\++\order{\S^{-1/2}},\\
\label{eq_sup:HP_S^+_SWA}
\hat{s}_{\r_i\alpha}^+&=\sqrt{2\S}\hat{b}_{\r_i\alpha}+\order{\S^{-1/2}},\\
\label{eq_sup:HP_S^z_SWA}
\hat{s}_{\r_i\alpha}^z&=\S-\hat{b}_{\r_i\alpha}^\+\hat{b}_{\r_i\alpha}.
\end{align}
In the spin-wave approximation, $\order{\S^{-1/2}}$ terms corresponding to higher-order contributions to spin fluctuations are neglected.
Using these expansions and $\hat{b}_{\r_i\alpha}=\frac{1}{\sqrt{\N}}\sum_\k e^{i\k\cdot\r_i}\hat{b}_{\k\alpha}$, the Heisenberg Hamiltonian $\hat{H}_\alpha$ can be diagonalized as
\begin{align}\label{eq_sup:H_SWA}
\hat{H}_\alpha&=\sum_\k E_{\k\alpha}\hat{b}_{\k\alpha}^\+\hat{b}_{\k\alpha}-3J\,\S^2\N-h_\alpha\S\N+\order{\S^{-1}},
\end{align}
where the quasiparticle described by $\hat{b}_{\k\alpha}^\+$ and $\hat{b}_{\k\alpha}$ is called magnon and exhibits the energy spectrum $E_{\k\alpha}=h_\alpha+2J\S\qty(3-\sum_{\mu=x,y,z}\cos(k_\mu))$.
In particular, a magnon with small $\k$ obeys the parabolic dispersion relation
\begin{align}\label{eq_sup:E_k}
E_{\k\alpha}\simeq h_\alpha+J\S\k^2.
\end{align}
Since we are interested in the low-temperature regime ($T_\alpha\ll J$), we hereafter use this parabolic form for magnons at low energy.

Next, we evaluate the tunneling currents within the spin-wave approximation.
By substituting Eqs.~\eqref{eq_sup:HP_S^-_SWA}--\eqref{eq_sup:HP_S^z_SWA} into Eqs.~\eqref{eq_sup:G^+-_ret} and \eqref{eq_sup:G^zz_ret}, the retarded response functions in the spin-wave approximation are found to be
\begin{align}\label{eq_sup:G^+-_ret_SWA}
G_{\k\alpha}^{+-;\mathrm{ret}}(\omega)&=\frac{2\S}{\omega^+-E_{\k\alpha}}+\order{(1/\S)^0},\\
\label{eq_sup:G^zz_ret_SWA}
G_{\k\alpha}^{zz;\mathrm{ret}}(\omega)&=\order{(1/\S)^0}.
\end{align}
From Eqs.~\eqref{eq_sup:I_S_1}--\eqref{eq_sup:G_zz(omega)}, \eqref{eq_sup:G^+-_ret_SWA}, and \eqref{eq_sup:G^zz_ret_SWA}, the spin and heat currents up to $\order{\S^2}$ are written as Eqs.~\eqref{eq:I_S_2} and \eqref{eq:I_H_2} in the main text, respectively:
\begin{align}\label{eq_sup:I_S_2}
I_\mathrm{S}&=\int_{-\infty}^\infty\!\!\dd{\omega}\T(\omega)\varDelta n_\mathrm{B}(\omega),\\
\label{eq_sup:I_H_2}
I_\mathrm{H}&=\int_{-\infty}^\infty\!\!\dd{\omega}(\omega+h_\mathrm{L})\T(\omega)\varDelta n_\mathrm{B}(\omega),\\
\T(\omega)&=\T_{+-}(\omega)=\sum_{\p\q}\frac{|\mathcal{J}_{\p\q}|^2}{2\pi}\A_{\p\mathrm{L}}(\omega+h_\mathrm{L})\A_{\q\mathrm{R}}(\omega+h_\mathrm{R}),
\end{align}
where $\A_{\k\alpha}(\omega)=-\Im\,G_{\k\alpha}^{+-;\mathrm{ret}}(\omega)=2\pi\S\delta(\omega-E_{\k\alpha})$ is the spectral function of magnons.
Within the spin-wave approximation, $\T_{zz}(\omega)$ arising from correlations of the $z$-component of spins do not contribute to $I_\mathrm{H}$.
Substituting $|\mathcal{J}_{\p\q}|=\frac{J_\mathrm{T}}{\N}$ and replacing the summation over momenta by integration, we can find that $\T(\omega)$ has the following simple form:
\begin{align}\label{eq_sup:G_+-(omega)_SWA}
\T(\omega)&=A\,\omega\,\theta(\omega).
\end{align}
In the $\S=1/2$ case realized with two-component bosonic atoms in an optical lattice, the factor is $A=\frac{(J_\mathrm{T})^2}{4\pi^{3}J^3}$.
Substituting Eq.~\eqref{eq_sup:G_+-(omega)_SWA} into Eqs.~\eqref{eq_sup:I_S_2} and \eqref{eq_sup:I_H_2} and using 
$(T_\alpha)^{d}F_d(x_\alpha)\equiv\int_0^\infty\!\!\dd{\omega}\frac{\omega^{d-1}}{\Gamma(d)}n_{\mathrm{B},\alpha}(\omega+h_\alpha)=(T_\alpha)^{d}\mathrm{Li}_d(e^{-x_\alpha})$, where $x_\alpha=h_\alpha/T_\alpha$, $\Gamma(d)$ is the Gamma function, and $\mathrm{Li}_d(z)=\sum_{k=1}^\infty z^k/k^d$ is the polylogarithm, 
we obtain
$I_\mathrm{S}$ and 
$I_\mathrm{H}$ expressed as Eqs.~\eqref{eq:I_S_SWA} and \eqref{eq:I_H_SWA} in the main text.

\section{Transport coefficients}
\begin{figure*}
    \centering
    \includegraphics[width=0.4\columnwidth,clip]{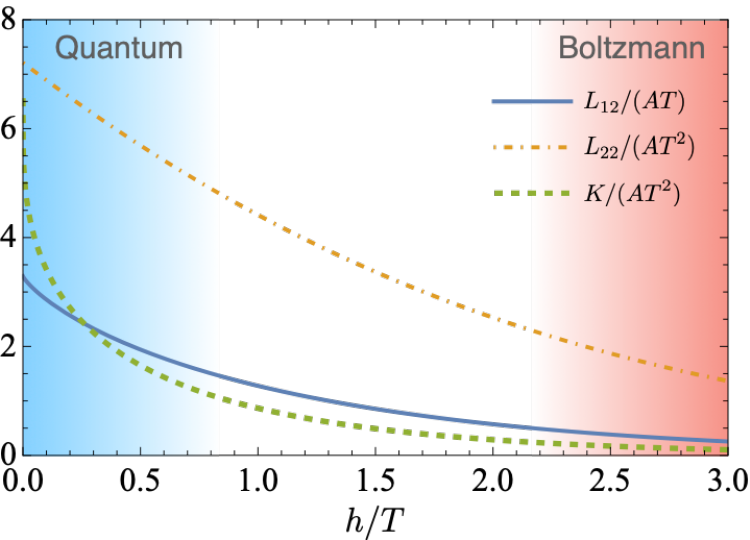}
    \caption{Transport coefficients as functions of the effective Zeeman field $h$.}
    \label{fig_sup:L_ij}
\end{figure*}

Here, we show numerical analysis and asymptotic behaviors of the Onsager coefficients $L_{ij}$  [Eqs.~\eqref{eq:L_11}--\eqref{eq:L_22} in the main text] and thermal conductance $K=L_{22}-T(L_{12})^2/L_{11}$.
Figure~\ref{fig_sup:L_ij} shows $L_{12}$, $L_{22}$, and $K$ as functions of $h/T$, while $L_{11}$ is plotted in Fig.~\ref{fig:transport}(a) in the main text.
We can see that as $h/T$ decreases all the quantities increases due to the quantum degeneracy.

In the strong field limit $x=h/T\to\infty$, thermal excitations in the FIs are regarded as the classical Boltzmann gases.
In this case, the transport coefficients 
become exponentially small as
\begin{align}
\frac{L_{11}}{AT}&=e^{-x}+\order{e^{-2x}},\\
\frac{L_{12}}{AT}&=(x+2)e^{-x}+\order{xe^{-2x}},\\
\frac{L_{22}}{AT^{2}}&=\qty(x^2+4x+6)e^{-x}+\order{x^2e^{-2x}},\\
\frac{K}{AT^{2}}&=2e^{-x}+\order{e^{-2x}}.
\end{align}
As a result, we obtain the Wiedemann-Frantz laws:
\begin{align}
\frac{L_{12}}{L_{11}}
=\frac{h}{T},\qquad
\frac{L_{22}}{L_{11}}
=\frac{h^2}{T},\qquad
\frac{K}{L_{11}}
=2T
\qquad(T/h\to+0).
\end{align}
The first two results have the same forms as in the case of the planar junction, while the last one is different by a factor $2$ from the planer case~\cite{nakata2018magnonic,Nakata2015-ns} due to the geometry of the junction.

In the quantum regime $x=h/T\to+0$, on the other hand, the transport coefficients, $L_{ij}$ and $K$, 
are enhanced by quantum degeneracy.
In addition, the coefficients except for $L_{22}$ exhibit nonanalytic behaviors because of the criticality at $h=0$ associated with the spontaneous magnetization.
Indeed, expanding Eqs.~\eqref{eq:L_11}--\eqref{eq:L_22} in the main text with respect to $x$ reads
\begin{align}
\frac{L_{11}}{AT}
&=-\ln x+\frac{x}{2}+\order{x^2},\\
\frac{L_{12}}{AT}
&=\frac{\pi^2}{3}+x(\ln x-2)+\order{x^3},\\
\frac{L_{22}}{AT^{2}}
&=6\zeta(3)-\frac{\pi^2 x}{3}+\order{x^2},\\
\frac{K}{AT^{2}}&=
    6\zeta(3)+\frac{\pi^4}{9\ln x}+\frac{\pi^2}{3}x+\order{\frac{x}{\ln x}}.
\end{align}
The spin conductance $L_{11}$ shows a divergent behavior.
While $L_{12}$ and $K$ are convergent at $x=0$, they nonanalytically approach to their values at $x=0$.
On the other hand, $L_{22}$ analytically behaves because the nonanalytic terms arising from the polylogarithm functions are completely canceled.

\section{Two-terminal relaxation dynamics}
This section is devoted to the details for the relaxation dynamics under the quasi-stationary assumption.
For convenience, we follow Refs.~\cite{brantut2013thermoelectric,grenier2012probing}.
To clarify the similarity of our two-terminal ferromagnetic system to two-terminal atomic gases~\cite{brantut2013thermoelectric,grenier2012probing}, we introduce the operator $\hat{N}_\alpha=\sum_{i}\hat{b}_{\r_i\alpha}^\+\hat{b}_{\r_i\alpha}=\S-\hat{M}_\alpha$ counting the number of bosons whose creation and annihilation operators are defined by the Holstein-Primakoff transformations [Eqs.~\eqref{eq_sup:HP_S^-} and \eqref{eq_sup:HP_S^+}].
For a while, we proceed without the use of the spin-wave approximation.
These bosons feel the effective Zeeman field $h_\alpha$ as a chemical potential $\mu_\alpha=-h_\alpha$.
In addition, we introduce the difference of the boson number $\Delta N=N_\rmL-N_\rmR$ and that of entropy $\Delta S=S_\rmL-S_\rmR$.
Note that $\Delta N$ is related to the magnetization difference $\varDelta M=M_\rmL-M_\rmR$ by $\varDelta M=-\varDelta N$.
We consider the situation where 
the spin and heat currents linearly respond to the biases [Eq.~\eqref{eq:linear-response} in the main text].
In terms of $\Delta N$ and $\Delta S$, the linear response is given by
\begin{align}\label{eq:current-bias3}
\dv{t}
\begin{pmatrix}
\varDelta N\\
\varDelta S
\end{pmatrix}
=-2
\begin{pmatrix}
I_{S}\\
I_\mathrm{H}/T
\end{pmatrix}
=-G
\begin{pmatrix}
1&\alpha_{ch}\\
\alpha_{ch}&L+\alpha_{ch}^2
\end{pmatrix}
\begin{pmatrix}
\varDelta \mu\\
\varDelta T
\end{pmatrix},
\end{align}
where $\varDelta \mu=\mu_\rmL-\mu_\rmR=\varDelta h$ and
\begin{align}\label{eq_sup:transport}
G=2L_{11},\qquad\alpha_{ch}=\frac{L_{12}}{L_{11}},\qquad 
L=\frac{L_{22}}{TL_{11}}-\qty(\frac{L_{12}}{L_{11}})^2.
\end{align}

Under the quasi-stationary condition where the thermalization within each FI much more rapidly occurs than the relaxation of $\varDelta N$ and $\varDelta T$ via an MQPC, $\varDelta N$ and $\varDelta S$ are related to the small biases by the following thermodynamic relation:
\begin{align}\label{eq_sup:thermodynamics}
    \begin{pmatrix}
    \varDelta N\\
    \varDelta S
    \end{pmatrix}
    =
    \begin{pmatrix}
    \qty(\pdv{N}{\mu})_T&\qty(\pdv{N}{T})_\mu\\
    \qty(\pdv{S}{\mu})_T&\qty(\pdv{S}{T})_\mu\\
    \end{pmatrix}
    \begin{pmatrix}
    \varDelta \mu\\
    \varDelta T
    \end{pmatrix}
    =\kappa
    \begin{pmatrix}
    1&\alpha_{r}\\
    \alpha_{r}&l+\alpha_{r}^2
    \end{pmatrix}
    \begin{pmatrix}
    \varDelta \mu\\
    \varDelta T
    \end{pmatrix},
\end{align}
where thermodynamic coefficients are given by
\begin{align}\label{eq_sup:thermodynamic_coefficients}
    \kappa=\qty(\pdv{N}{\mu})_T,\qquad
    \alpha_r=\qty(\pdv{S}{N})_T,\qquad
    l=\frac{C_{N}}{\kappa T}=\frac{1}{\kappa}\qty(\pdv{S}{T})_{N}.
\end{align}
In Eqs.~\eqref{eq_sup:thermodynamics} and \eqref{eq_sup:thermodynamic_coefficients}, thermodynamic quantities are evaluated with the averaged values: $N=(N_\rmL+N_\rmR)/2$, $S=(S_\rmL+S_\rmR)/2$, $\mu=(\mu_\rmL+\mu_\rmR)/2$.
In the context of bosons, $\kappa$ is the isothermal compressibility, $\alpha_r$ is the dilatation coefficient, and $C_{N}$ is the heat capacity with fixed $N$.
In the context of FIs, these quantities are written as $\kappa=\qty(\pdv{M}{h})_T
$, 
$\alpha_r=-\qty(\pdv{S}{M})_T
$, and
$l=\frac{1}{\kappa}\qty(\pdv{S}{T})_{M}
$ with $M=(M_\rmL+M_\rmR)/2$.

Combining Eq.~\eqref{eq:current-bias3} with Eq.~\eqref{eq_sup:thermodynamics}, we obtain the following differential equation governing the dynamics:
\begin{align}\label{eq_sup:relaxation}
   \tau_0\dv{t}
    \begin{pmatrix}
    \varDelta N/\kappa\\
    \varDelta T    
    \end{pmatrix}
    =-
    \begin{pmatrix}
    1&-\alpha\\
    -\frac{\alpha}{l}&\frac{L+\alpha^2}{l}
    \end{pmatrix}
    \begin{pmatrix}
    \varDelta N/\kappa\\
    \varDelta T  
    \end{pmatrix},
\end{align}
where
\begin{align}\label{eq_sup:tau_0}
\tau_0=\frac{\kappa}{2L_{11}},\qquad
\alpha=\alpha_r-\alpha_{ch}.
\end{align}
The solution of Eq.~\eqref{eq_sup:relaxation} is given by
\begin{align}\label{eq_sup:Delta_N,T(t)}
\begin{pmatrix}
\varDelta N(t)/\N\\
\varDelta T(t)/T
\end{pmatrix}
=
\begin{pmatrix}
\tilde{\Lambda}_{11}(t)&\tilde{\Lambda}_{12}(t)\\
\tilde{\Lambda}_{21}(t)&\tilde{\Lambda}_{22}(t)
\end{pmatrix}
\begin{pmatrix}
\varDelta N(0)/\N\\
\varDelta T(0)/T
\end{pmatrix},
\end{align}
where
\begin{align}\label{eq_sup:Lambda_11}
\tilde{\Lambda}_{11}(t)&=\frac{1}{2}\qty{\qty(e^{-t/\tau_-}+e^{-t/\tau_+})+\frac{\frac{L+\alpha^2}{l}-1}{\lambda_+-\lambda_-}\qty(e^{-t/\tau_-}-e^{-t/\tau_+})},\\
\label{eq_sup:Lambda_12}
\tilde{\Lambda}_{12}(t)&=\qty(\frac{T\kappa}{\N})^2l\tilde{\Lambda}_{21}(t)=\frac{T}{\N}\frac{\alpha \kappa}{\lambda_+-\lambda_-}\qty(e^{-t/\tau_-}-e^{-t/\tau_+}),\\
\label{eq_sup:Lambda_22}
\tilde{\Lambda}_{22}(t)&=\frac{1}{2}\qty{\qty(e^{-t/\tau_-}+e^{-t/\tau_+})-\frac{\frac{L+\alpha^2}{l}-1}{\lambda_+-\lambda_-}\qty(e^{-t/\tau_-}-e^{-t/\tau_+})}.
\end{align}
The relaxation times $\tau_\pm = \tau_0/\lambda_\pm$ are characterized by
\begin{align}\label{eq_sup:lambda_pm}
\lambda_\pm=\frac{1}{2}\qty(1+\frac{L+\alpha^2}{l})\pm\sqrt{\frac{\alpha^2}{l}+\frac{1}{4}\qty(1-\frac{L+\alpha^2}{l})^2},
\end{align}
which are eigenvalues of the matrix in Eq.~\eqref{eq_sup:relaxation}.
Note that the signs before $\frac{\frac{L+\alpha^2}{l}-1}{\lambda_+-\lambda_-}$ in Eqs.~\eqref{eq_sup:Lambda_11} and \eqref{eq_sup:Lambda_22} are consistent with those in Ref.~\cite{grenier2012probing} but opposite to those in Ref.~\cite{brantut2013thermoelectric}.
In terms of spin systems, Eq.~\eqref{eq_sup:Delta_N,T(t)} is rewritten as
\begin{align}\label{eq_sup:Delta_M,T(t)}
\begin{pmatrix}
\varDelta M(t)/\N\\
\varDelta T(t)/T
\end{pmatrix}
=
\begin{pmatrix}
\Lambda_{11}(t)&\Lambda_{12}(t)\\
\Lambda_{21}(t)&\Lambda_{22}(t)
\end{pmatrix}
\begin{pmatrix}
\varDelta M(0)/\N\\
\varDelta T(0)/T
\end{pmatrix}
,\qquad
\begin{pmatrix}
\Lambda_{11}(t)&\Lambda_{12}(t)\\
\Lambda_{21}(t)&\Lambda_{22}(t)
\end{pmatrix}
\equiv
\begin{pmatrix}
\tilde{\Lambda}_{11}(t)&-\tilde{\Lambda}_{12}(t)\\
-\tilde{\Lambda}_{21}(t)&\tilde{\Lambda}_{22}(t)
\end{pmatrix}.
\end{align}
We note that, as long as the biases are small and the quasi-stationary condition is valid, this solution for relaxation dynamics holds for any $h$ and $T$ and applicable beyond the situation where the spin wave approximation is justified.

We now focus on the highly spin polarized case with $T\ll J$, where the spin-wave approximation is applicable.
Substituting Eqs.~\eqref{eq:L_11}--\eqref{eq:L_22} in the main text into $\alpha_{ch}$ and $L$ in Eq.~\eqref{eq_sup:transport} yields
\begin{align}\label{eq_sup:transport_SWA}
\alpha_{ch}=\frac{2F_{2}(x)}{F_{1}(x)}+x,\qquad
L=\frac{6F_{3}(x)}{F_{1}(x)}-4\qty(\frac{F_{2}(x)}{F_{1}(x)})^2.
\end{align}
On the other hand, $\kappa$, $\alpha_r$, and $l$ in Eq.~\eqref{eq_sup:thermodynamic_coefficients} within the spin-wave approximation can be evaluated similarly to the calculation of thermodynamic quantities for an ideal Bose gas, and are determined as follows:
\begin{align}\label{eq_sup:thermodynamics_SWA}
    \frac{\kappa}{B\sqrt{T}}=F_{\frac{1}{2}}(x),\qquad
    \alpha_r=\frac{3}{2}\frac{F_{\frac{3}{2}}(x)}{F_{\frac{1}{2}}(x)}+x,\qquad
    l=\frac{15}{4}\frac{F_{\frac{5}{2}}(x)}{F_{\frac{1}{2}}(x)}-\frac{9}{4}\qty(\frac{F_{\frac{3}{2}}(x)}{F_{\frac{1}{2}}(x)})^2,
\end{align}
where $B=\N/(4\pi J\S)^{3/2}=\N/(2\pi J)^{3/2}$ for $\S=1/2$.
Finally, $\tau_0$ and $\alpha$ in Eq.~\eqref{eq_sup:tau_0} are found to be
\begin{align}\label{eq_sup:tau_0-alpha_SWA}
\tau_0=\frac{F_{\frac{1}{2}}(x)}{F_{1}(x)}\tilde{\tau},
\qquad
\alpha=\frac{3F_{\frac{3}{2}}(x)}{2F_{\frac{1}{2}}(x)}-\frac{2F_{2}(x)}{F_{1}(x)}
\end{align}
with $\tilde{\tau}=B/(2A\sqrt{T})$.

We now focus on the quantum regime at a weak effective Zeeman fields $x=h/T\ll1$.
Expanding prefactors in $\tilde{\Lambda}_{ij}$ [Eqs.~\eqref{eq_sup:Lambda_11}--\eqref{eq_sup:Lambda_22}] with respect to $x$, we obtain
\begin{align}
\Lambda_{11}(t)&\simeq e^{-t/\tau_-},\\
\Lambda_{12}(t)&=\order{(T/J)^{3/2}},\\
\Lambda_{21}(t)&\simeq\frac{\sqrt{2} \pi ^3  }{9 \zeta (3)}\sqrt{\frac{x}{(T/J)^{3}}}\qty(e^{-t/\tau_-}-e^{-t/\tau_+}),\\
\Lambda_{22}(t)&\simeq e^{-t/\tau_+}.
\end{align}
Since $T/J\ll1$ is now considered, $\Lambda_{12}(t)$ is negligible.
For sufficiently small $x\ll(T/J)^3$, $\Lambda_{21}(t)$ is also vanishingly small.
Therefore, the diagonal terms only survive in the limit of $x=h/T\to+0$, leading to
\begin{align}
\begin{pmatrix}
\varDelta M(t)\\
\varDelta T(t)
\end{pmatrix}
\simeq
\begin{pmatrix}
e^{-t/\tau_-}&0\\
0&e^{-t/\tau_+}
\end{pmatrix}
\begin{pmatrix}
\varDelta M(0)\\
\varDelta T(0)
\end{pmatrix}.
\end{align}
This means that time evolution of  $\varDelta M(t)$ is decoupled from that of $\varDelta T(t)$.
We note that, in the regime $(T/J)^3\ll x\ll1$, the magnitude of $\Lambda_{21}(t)$ becomes larger than those of $\Lambda_{11}(t)$ and $\Lambda_{22}(t)$.

Next, we evaluate the relaxation times $\tau_\pm = \tau_0/\lambda_\pm$.
From Eqs.~\eqref{eq_sup:lambda_pm}--\eqref{eq_sup:tau_0-alpha_SWA}, the relaxation times $\tau_\pm$ for $x\to+0$ are found to be
\begin{align}
\tau_+&=\frac{5\zeta \left(\frac{5}{2}\right)}{8\zeta (3)}\tilde{\tau}=0.697496\times\tilde{\tau},\\
\tau_-&=\frac{\sqrt{\pi}}{-\sqrt{x}\ln x}\tilde{\tau}\gg\tilde{\tau}.
\end{align}
Therefore, in the case of $x\ll(T/J)^3\ll1$, the decay time $\tau_-$ of $\varDelta M(t)$ diverges, so that $\varDelta M(t)$ decays much more slowly than $\varDelta T(t)$ with its decay time $\tau_+$. 
In the Boltzmann limit ($h/T\to\infty$), on the other hand, the relaxation times become comparable to $\tilde{\tau}$ as
\begin{align}
\tau_+&=\frac{12\tilde{\tau}}{15+\sqrt{33}}=0.578465\tilde{\tau},\\
\tau_-&=\frac{12\tilde{\tau}}{15-\sqrt{33}}=1.29654\tilde{\tau}.
\end{align}

\section{Estimation of relaxation time scale in a 
typical experimental setup}
Here, we show that the relaxation dynamics discussed throughout this Letter is accessible in current cold-atom experiments by estimating $\tilde{\tau}=\hbar\sqrt{\pi^3/2}\mathcal{N}(J/J_\mathrm{T})^2/\sqrt{k_\mathrm{B}JT}$ for a typical experimental setup.
We obtain the time scale $\tilde{\tau} = 1.74\,\mathrm{sec}$ by using the following values of the parameters:
$J_\mathrm{T} / J = 0.5$, $k_\mathrm{B}T / J = 0.7$, $\mathcal{N} = 54$, and $J / \hbar = 93 \times 2\pi\, \mathrm{Hz}$.
We expect that this temperature scale is accessible as in experiments of Fermi Hubbard systems ~\cite{Chiu2019-up,Shao2024-uc}, which is usually difficult to achieve a low-temperature regime
compared to the bosonic counterpart.
The value of $J$ is taken from the experimental value in Ref.~\cite{Jepsen2020-fg}.
Note that the value of $J$ in Ref.~\cite{Jepsen2020-fg} is scaled by Planck constant $2\pi\hbar$, while our $J$ is scaled by $\hbar$.\\

\section{Finite-size effect}
We consider that the effect of a finite size in experiments is not so relevant if we take a relatively large system size, for example, $\mathcal{N} > 50$. According to Euler–Maclaurin formula, the corrections $\sim O(1/\mathcal{N}^2)$ coming from a finite system size $\mathcal{N}$ to currents in Eqs.~\eqref{eq:I_S_SWA} and \eqref{eq:I_H_SWA} arise when replacing summation over momentum in $\mathcal{T}(\omega)$ with integral. Therefore,  the correction may be negligible for $\mathcal{N} > 50$.

\section{Control of effective Zeeman fields in cold-atom experiments}
Here, we discuss how to experimentally control the effective Zeeman fields $h_\alpha$ for FIs in thermal equilibrium, i.e, the initial state for relaxation dynamics.
We start by reviewing the setup to realize ferromagnetic Heisenberg model in cold-atom experiments.
For quantum simulation of Heisenberg spins a deep Mott phase of two-component bosons is utilized.
We refer to these two internal states as $\sigma=\uparrow,\downarrow$, with the number $N_{\sigma,\alpha}$
of atoms in each state.
The particle number difference between the two states $N_{\uparrow,\alpha}-N_{\downarrow,\alpha}$, is assigned to the $z$-component of the magnetization $M_\alpha=(N_{\uparrow,\alpha}-N_{\downarrow,\alpha})/2$.
The thermodynamic quantity $h_\alpha$ conjugate to $M_\alpha$ is given by $h_\alpha=\qty(\pdv{\mathcal{E}_\alpha}{M_\alpha})_{S_\alpha}$, where $\mathcal{E}_\alpha$ and $S_\alpha$ are the internal energy and entropy, respectively.
In the absence of external fields, the particle numbers $N_{\sigma,\alpha}$ given by initial state preparation determine $M_\alpha$.
Therefore, the magnetization is directly controllable in the cold-atom experiments.
On the other hand, $h_\alpha$ at fixed temperature is given as a Lagrange multiplier to fix the magnetization as in the case of Fermi gases~\cite{Krinner2016-gd}.
In this sense, $h_\alpha$ is not an external field but an effective one.
The value of $h_\alpha$ can be experimentally evaluated by measuring the internal energy~\cite{Takasu2020Energy}.
We note that this situation is different from that for solid FIs, where $h_\alpha$ is given as an external magnetic field.
Although the parameters directly controlled in the experiments vary between cold atoms and solid FIs, both are governed by the same Heisenberg model and share the same magnonic ciriticality.

We next turn to a scheme to realize thermal initial states in the quantum regime ($h_\alpha/T_\alpha\ll1$) near magnonic critical point.
To obtain sufficiently weak fields $h_\alpha$, one has to fine-tune the filling factor $\nu_{\downarrow,\alpha}=N_{\downarrow,\alpha}/\N$ in the highly polarized regime $\nu_{\downarrow,\alpha}\ll1$ as theoretically estimated in the last part of this section.
(Note that $M_\alpha$ is related to the filling by $M_\alpha/\N=1/2-\nu_\alpha$.)
Here, we explain a way to control the filling factors $\nu_{\downarrow,\alpha}$ of two FIs independently with accuracy $\Delta \nu_{\downarrow,\alpha}\sim0.01$ by using Rabi oscillation of one reservoir with the other reservoir being off-resonant due to a light shift.
The detailed steps in the scheme are as follows:

\begin{enumerate}
    \item Prepare a deep optical lattice, in which all lattice sites are spatially decoupled from each other, and load one spin-up atom per lattice site.
    \item Induce a light shift of the down state on atoms in the right reservoir by using a digital mirror devise.
    \item Apply microwaves with frequency $\omega=\omega_{\downarrow,\mathrm{L}}-\omega_{\uparrow,\mathrm{L}}$ to the system in order to induce the Rabi oscillation between spin-up and spin-down levels for atoms in the left reservoir.
    The system is irradiated until $\nu_{\downarrow,\mathrm{L}}$ reaches the target value.
    Due to the light shift, atoms in the right reservoir are off-resonant ($\omega\neq\omega_{\downarrow,\mathrm{R}}-\omega_{\uparrow,\mathrm{R}}$) and remain spin-up.
    \item Remove the light shift for the right reservoir and apply a light shift for the left one.
    \item Apply microwaves with frequency $\omega=\omega_{\downarrow,\mathrm{R}}-\omega_{\uparrow,\mathrm{R}}$ for atoms in the right reservoir to achieve the target value of $\nu_{\downarrow,\mathrm{R}}$ via the Rabi oscillation.
    Due to the light shift, atoms in the left reservoir are off-resonant and thus $\nu_{\downarrow,\mathrm{L}}$ remains unchanged.
    \item Remove the light shift and make the lattice potentials of both reservoirs shallow with keeping the channel close.
    \item Wait until both reservoirs are thermalized.
\end{enumerate}

\begin{figure*}[h]
    \centering
    \includegraphics[width=0.4\columnwidth,clip]{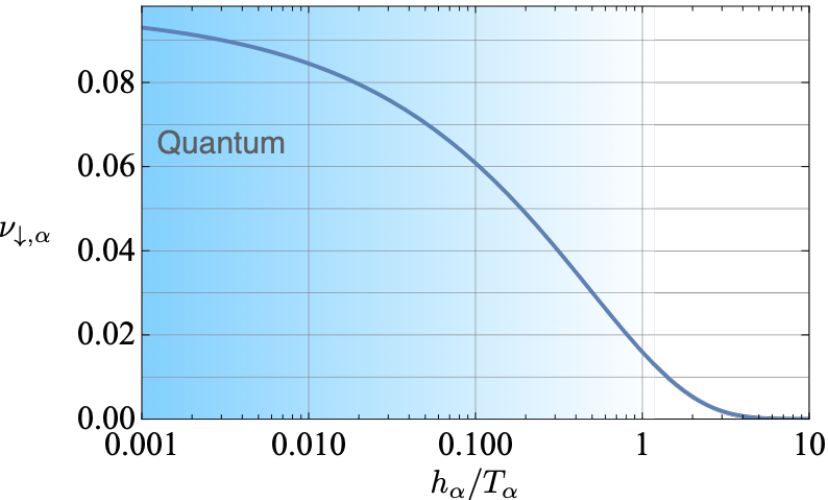}    
    \caption{Filling factor $\nu_{\downarrow,\alpha}$ at $T_\alpha=0.7J$ as a function of $h_\alpha/T_\alpha$.
    The quantum regime (blue region with $h_\alpha/ T_\alpha\ll1$) can be realized by tuning $\nu_{\downarrow,\alpha}$.}
    \label{fig_sup:nu_down}
\end{figure*}

To theoretically estimate the filling factor $\nu_{\downarrow,\alpha}=N_{\downarrow,\alpha}/\N$ for the quantum regime, we employ the spin-wave approximation.
By using $M_\alpha=(N_{\uparrow,\alpha}-N_{\downarrow,\alpha})/2=\frac{1}{2}\N-\sum_\k n_{\mathrm{B},\alpha}(E_{\k\alpha})$ and  $\N=N_\uparrow+N_\downarrow$ for the Mott phase, the filling factor of the down component reads
\begin{align}\label{sup_eq:N_down}
    \nu_{\downarrow,\alpha}
    =\frac{1}{\N}\sum_\k n_{\mathrm{B},\alpha}(E_{\k\alpha})
    =\qty(\frac{T_\alpha}{2\pi J})^{3/2}F_{3/2}(h_\alpha/T_\alpha).
\end{align}
In Fig.~\ref{fig_sup:nu_down}, the filling factor $\nu_{\downarrow,\alpha}$ at $T_\alpha =0.7J$ with the spin-wave approximation is plotted as a function of $h_\alpha/T_\alpha$. This figure shows that a small effective magnetic field in the quantum regime (blue region) can be obtained by changing the filling factor $\nu_{\downarrow,\alpha}$ with accuracy $\Delta \nu_{\downarrow,\alpha}=0.01$.
We consider that the realization of ferromagnetic systems at $T_\alpha\sim0.7J$ is feasible because this temperature scale has been achieved in experiments with optical lattice~\cite{Chiu2019-up,Shao2024-uc}.
In addition, we note that our analysis with the spin-wave approximation is valid because $T_\alpha =0.7J$ is much lower than the Curie temperature ($T_\mathrm{c}=6J$ in the mean-field theory) and the reservoirs show sufficiently large magnetization $M_\alpha/(\frac{1}{2}\mathcal{N_\alpha})\sim 0.8$.

We finally mention that if different external fields are applied to two reservoirs instead of the effective Zeeman fields tuned by filling factors, the resulting critical dynamics is altered from that focused on throughout this Letter.
In our work, we focus on the situation where two FIs experience effective Zeeman fields appearing in the thermal density operator in Eq.~\eqref{eq_sup:rho_H} tuned by the initial filling fractions (or magnetizations $M_\alpha$).
In this case, both FIs have the same magnon DoS, $\rho_{\alpha=\mathrm{L,R}}(\omega)=\sum_{\q}\A_{\q\alpha}(\omega+h_\alpha)\propto\sqrt{\omega}$, and the resulting transmittance $\mathcal{T}(\omega)\propto\rho_{\mathrm{L}}(\omega)\rho_{\mathrm{R}}(\omega)\propto\omega$ causes anomalous transport in the quantum regime.
On the other hand, when left and right FIs are subjected to actual fields, the difference in the fields shifts the magnon energy level in the left FI from that in the right FI, leading to $\rho_{\mathrm{L}}(\omega)\neq\rho_{\mathrm{R}}(\omega)$.
As a result, the dynamics in the quantum regime become different from those in our case with effective Zeeman fields.
In addition, the energy shift by different actual fields may suppress the spin-exchange processes between left and right FIs.

\section{Effect of number of spin pairs linking reservoirs}
\begin{figure*}[h]
    \centering
    \includegraphics[width=0.4\columnwidth,clip]{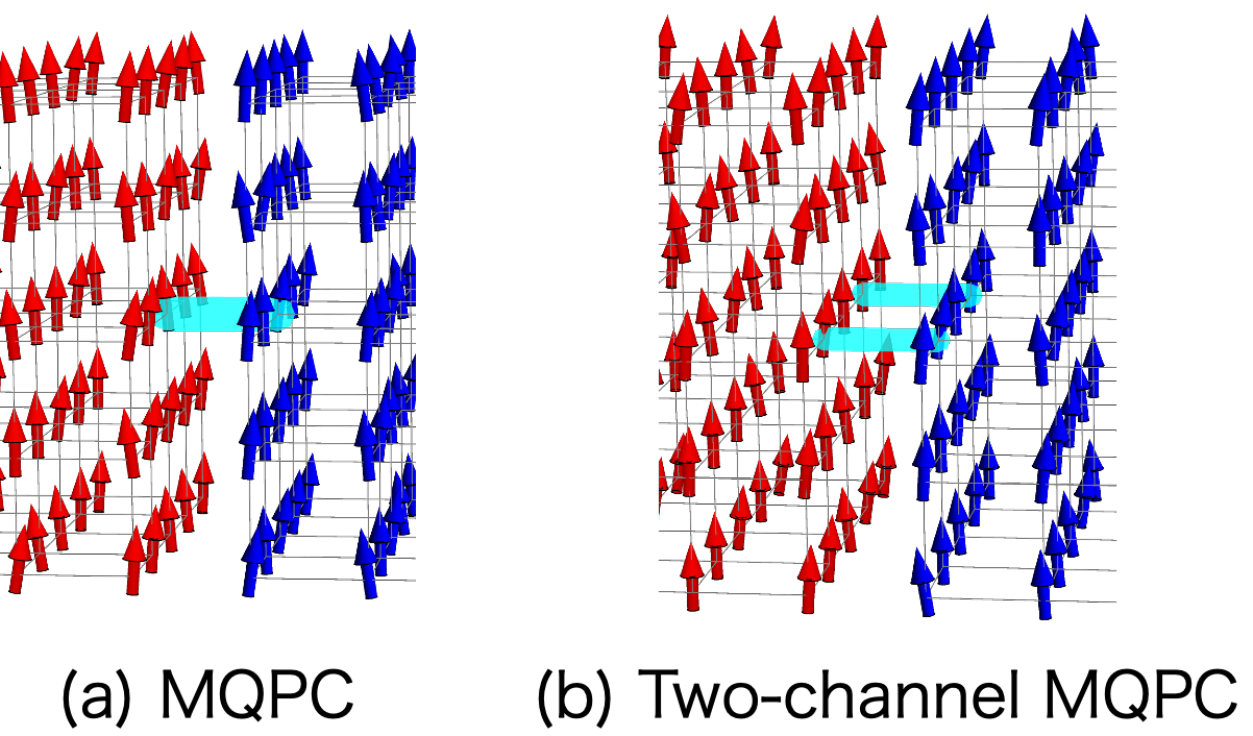}    
    \caption{
    Junctions with different number of spin pairs (light blue bonds) connecting FIs. (a) Single-channel MQPC with a single pair (b) Two-channel MQPC with two pairs spatially separated from each other.}
    \label{fig_sup:multichannel}
\end{figure*}

Here, we discuss how the number of spins connecting FIs affects the tunneling currents.
When there are a few spin pairs on the interface as in Fig.~\ref{fig_sup:multichannel} (b), the transport may be qualitatively similar to that in the MQPC case in Fig.~\ref{fig_sup:multichannel} (a). 
In the situation for Fig.~\ref{fig_sup:multichannel} (b), the spin correlations between the spatially separated pairs are almost negligible and thus the links can be regarded as independent MQPCs.
Then, the resulting current $I_\mathrm{S/H}\simeq\N_\mathrm{ch}I_\mathrm{S/H}^{\mathrm{(MQPC)}}$ would be given by that for the single MQPC value $I_\mathrm{S/H}^{\mathrm{(MQPC)}}$ [Eqs.~\eqref{eq:I_S_SWA} and \eqref{eq:I_H_SWA} in the main text] multiplied by the number of pairs $\N_\mathrm{ch}$.
Therefore, the magnonic criticality causes the divergence in the spin conductance and slowing down in spin relaxation as in the MQPC case.

\section{Similarity to a two-terminal system of normal Bose gases}
In this section, we discuss that weakly interacting Bose gases linked with each other via a quantum point contact is expected to exhibits anomalous transport by criticality as FIs connected via MQPC.
It is known that, in highly polarized FIs where the spin-wave approximation is applicable, the thermally excited magnons under effective magnetic fields $h_\alpha$ can be regarded as noninteracting Bose gases with the chemical potentials $\mu_\alpha = -h_\alpha$~\cite{takahashi1986quantum}.
This relates the magnonic critical point ($h_\alpha\to+0$)  to the Bose-Einstein transition point ($\mu_\alpha\to-0$).
From this correspondence, our results suggest that the bosonic two-terminal system exhibits divergence in mass conductance and very slow relaxation of particle number difference when both Bose gases are near the Bose-Einstein transition points.
Such anomalous transport and relaxation by the criticality may survive in the limit of a weak interaction for bosons.

\end{document}